\documentclass[conference]{IEEEtran}
\IEEEoverridecommandlockouts

\usepackage{amssymb}
\usepackage{graphicx}
\usepackage{url}
\usepackage{float}
\usepackage{amsfonts}
\usepackage{amsmath}
\usepackage{nccmath}
\usepackage{adjustbox}
\usepackage{multirow}
\usepackage{eurosym}
\usepackage{mathtools}
\usepackage{textcomp}
\usepackage{scrextend}
\usepackage{verbatim}
\usepackage{soul} 
\usepackage{fancyhdr,lipsum}
\usepackage{enumitem}

\usepackage[utf8]{inputenc}
\usepackage{amsmath}
\usepackage{amsfonts}
\usepackage{mathtools}
\usepackage{physics}
\usepackage{dsfont}
\usepackage{verbatim}
\usepackage{color}
\usepackage[dvipsnames]{xcolor}
\usepackage{soul}
\usepackage[utf8]{inputenc}
\usepackage[ruled,longend]{algorithm2e}
\usepackage{textgreek}
\usepackage{amssymb}
\usepackage{multirow}

\usepackage{tcolorbox}

\usepackage{balance}
\usepackage{graphicx}
\usepackage{subfigure}
\usepackage{url}
\usepackage{nccmath}
\usepackage{adjustbox}
\usepackage{multirow}
\usepackage{eurosym}

\usepackage{cite}
\usepackage{amsmath,amssymb,amsfonts}
\usepackage{algorithmic}
\usepackage{graphicx}
\usepackage{textcomp}
\usepackage{xcolor}
\def\BibTeX{{\rm B\kern-.05em{\sc i\kern-.025em b}\kern-.08em
    T\kern-.1667em\lower.7ex\hbox{E}\kern-.125emX}}
    
\begin{document}

\title{Quantum Computing Approach for Energy Optimization in a Prosumer Community
}

\author{\IEEEauthorblockN{Carlo Mastroianni}
\IEEEauthorblockA{\textit{ICAR-CNR}\\ Via P. Bucci, 8/9 C, Rende, Italy \\
carlo.mastroianni@icar.cnr.it}
\and
\IEEEauthorblockN{Luigi Scarcello}
\IEEEauthorblockA{\textit{ICAR-CNR}\\ Via P. Bucci, 8/9 C, Rende, Italy \\
luigi.scarcello@icar.cnr.it}
\and
\IEEEauthorblockN{Jacopo Settino}
\IEEEauthorblockA{\textit{Dipartimento di Fisica, Università della Calabria}\\
Via P. Bucci, Rende, Italy \\
\textit{ICAR-CNR}, Via P. Bucci, 8/9 C, Rende, Italy \\
\textit{INFN, gruppo collegato di Cosenza}, Rende, Italy \\
jacopo.settino@unical.it}
}

\maketitle

\begin{abstract}
This paper presents a quantum approach for the formulation and solution of the prosumer problem, i.e., the problem of minimizing the energy cost incurred by a number of users in an energy community, while addressing the constraints given by the balance of energy and the user requirements. As the problem is NP-complete, a hybrid quantum/classical algorithm could help to acquire a significant speedup, which is particularly useful when the problem size is large. This work describes the steps through which the problem can be transformed, reformulated and given as an input to Quantum Approximate Optimization Algorithm (QAOA), and reports some experimental results, in terms of the quality of the solution and time to achieve it, obtained with a quantum simulator, when varying the number of constraints and, correspondingly, the number of qubits.
\end{abstract}

\section{Introduction}
\label{secIntro}

Energy and climate policies are some of the greatest challenges in the modern society and a transition is required from exhaustible energy sources, such as fossil and nuclear, to more sustainable modes of production and consumption. Several strategies have been devised to support the employment of renewable energy sources (RES) and the development of energy communities \cite{nafi2016survey, giordano2020optimization}. An energy transition is needed not only in terms of energy sources but also towards a more local, decentralized and pervasive management of the energy. 
Through two European directives that are part of the Clean Energy for All Europeans Package, namely the Renewable Directive (2018/2001/EU) and the Electricity Market Directive (2019/944/EU), important innovations have been introduced, which involve citizens in the smarter exploitation of renewable sources.
The  Italian government, by approving the Milleproroghe Decree
(D.L. 162/2019)
has recently laid concrete foundations for the implementation of local energy communities, to foster collective self-consumption and energy exchange among prosumers.

Prosumers are community members that can both produce and consume the energy produced locally, and exchange the surplus energy directly within the community \cite{zafar2018prosumer}.
The surplus of energy from RES is traded to meet the local energy demand, thus allowing a significant reduction of energy consumption and energy costs.
The process of energy sharing among prosumers involves several optimization techniques, in which, typically, a central coordinator
performs a virtual aggregation of the prosumers in groups and orchestrates the optimization process \cite{scarcello2022cascade, scarcello2022edge}.
In most cases, price alerts are used to negotiate the energy exchanges among individual market participants with the aim to drive prosumers in the management of their RES plants and electrical devices \cite{liu2017energy}.

Recently, research and industrial efforts are showing that quantum computing can offer the opportunity to approach energy optimization problems with a completely different paradigm and with significant potential advantages in the medium and long term \cite{QCForEnergySystems}. Variational algorithms have been proposed in which the computation is hybrid: (i) the most intensive computation is done on quantum hardware, and is driven by a number of tunable parameters, while (ii) the optimization of these parameters is performed on a classical computer, for example with techniques similar to gradient descent \cite{biamonte2017, dunjko2016}. While this research field is still in its infancy, there is great hope that the intrinsic quantum parallelism can help to speed up the exploration of the search space, that is almost always exponential in optimization problems regarding the consumption and sharing of energy. 

In this paper, we show how the prosumer problem -- i.e., the problem of optimizing the scheduling of electrical loads, and minimizing the related cost, given the constraints related to the power grid and user requirements -- can be formulated and given as an input to one of the most renowned hybrid quantum/classical algorithms, i.e., the Quantum Approximate Optimization Algorithm (QAOA) \cite{QAOA}. In a nutshell, the prosumer problem is first formulated as an Integer Linear Programming (ILP), then, with some intermediate steps, it is transformed into a problem that can be tackled by QAOA, i.e., in physical terms, finding the value of minimum energy of a Hamiltonian operator. This corresponds to finding the minimum eigenvalue of a diagonal operator: the eigenvector associated with that eigenvalue represents the state of minimum energy -- i.e., the \textit{ground state} -- of the Hamiltonian and, at the same time, codifies, as a string of binary variables, the solution of the optimization problem.


The steps can be summarized as follows:
\begin{enumerate}
    \item first, the prosumer problem is formulated as an ILP problem, where the objective is to minimize the energy cost while satisfying a number of constraints related to the maximum available energy and user requirements;
    \item the problem is transformed into a Quadratic Unconstrained Binary Optimization (QUBO) problem \cite{qubo2007}, where the objective function contains binary variables in a linear or quadratic form and the constraints are removed and incorporated into the objective function;
    \item substitution of variables is operated to convert the problem into an Ising problem \cite{ising2000}, where the values $\{0,1\}$ of the binary variables are associated, respectively, with values $\{1,-1\}$ of discrete variables;
    \item the Ising problem is formulated as a Hamiltonian operator, which is used to measure the energy of a number of qubits, where each qubit corresponds to one discrete variable of the Ising problem;
    \item the QAOA is used to find the minimum value of energy, i.e., the minimum eigenvalue, of the Hamiltonian operator. The value of energy is obtained by composing the results, $1$ or $-1$, of the measurements of single qubits (see step 4) at the end of the QAOA execution. When the algorithm converges to the minimum eigenvalue, the values $\{1,-1\}$ are interpreted as the values $\{0,1\}$ of the QUBO problem, and therefore of the original prosumer problem.
\end{enumerate}

Given the multidisciplinary nature of this work, it is useful to specify the contribution of this paper, and also what this paper does not cover. The contribution is twofold:
\begin{enumerate}
    \item starting from a typical prosumer problem, we describe how the problem is transformed, step-by-step, to the problem of finding the ground state of a Hamiltonian operator. We explain how the prosumer problem can be transformed to be given as an input to QAOA, and how the QAOA result can be interpreted in terms of the prosumer problem. We hope that this can stimulate other researchers to investigate the opportunity of adopting quantum computation to approach the prosumer problem, and energy optimization problems in general;
    \item we report the results of some experiments for the small-size problem, which can be solved using the open-access simulator provided by the IBM Quantum Experience portal\footnote{https://quantum-computing.ibm.com/}. The size of the problem is limited by the need for simulating a quantum process with classical software and hardware. To solve larger problems, it is needed to resort to real quantum computers, which is the objective of current and future work.
\end{enumerate}

Instead, in this paper we do not provide:
\begin{enumerate}
    \item a general introduction to quantum computing basics. It can be found in excellent tutorial papers \cite{QuantumIntroNannicini, QuantumIntroRieffel}. In particular, in \cite{Coles2022QuantumAI} the reader can also find the implementation of many algorithms on the IBM platform, the same adopted in this paper;
    \item a complete description of the QAOA algorithm, in particular the process executed to find the ground state of the Hamiltonian operator. This description can be found on several papers, among which the original paper \cite{QAOA} and a recent paper that examines and compares the optimization algorithms that can be exploited in QAOA \cite{QAOA-Combarro}.
\end{enumerate}

The rest of the paper is organized as follows: Section  \ref{secILP} describes how a binary optimization problem can be transformed into the problem of finding the minimum eigenvalue of an operator, and therefore can be solved with the QAOA algorithm; Section \ref{secModel} describes how this procedure is specialized to the case of the prosumer problem, where each schedulable load is associated with a binary variable, which in turn is represented as a qubit; Section \ref{secExample} illustrates an example of a prosumer problem and reports the expression for the Hamiltonian operator that is given to the QAOA algorithm; Section \ref{secResults} reports a set of solutions obtained with the IBM quantum simulator; finally, Section \ref{secConclusions} concludes the paper.

\section{Solving a Binary Optimization Problem with QAOA}
\label{secILP}

As anticipated in the introductory section, this paper describes a general approach to transforming a prosumer problem into a problem that can be solved with the QAOA algorithm. The approach is partially inspired by the general procedure  presented in \cite{IsingNP} to handle a number of NP problems. However, in that work, the problem is not solved with QAOA, which can be run on a gate-based quantum computer, but with the quantum annealing approach, which is adopted by a completely different type of quantum computation, based on the adiabatic behaviour \cite{Annealing}.

This section focuses on the formal steps and aims to explain -- at least intuitively -- the reason why the QAOA algorithm can lead to the solution of a problem that is formulated as a binary optimization problem. The semantics of the expressions, in particular of the variables and constraints, will be given in Section \ref{secModel}.
The approach starts by defining the prosumer problem as a binary optimization problem, where the objective is to minimize the cost function $C$, expressed as in Eq. (\ref{eq:generalObjFunc}), and the constraints are in the form given in Eqs. (\ref{eq:generalConstraint}). 

\begin{equation}
\label{eq:generalObjFunc}
	C = \sum_{i=1}^{N}{c_{i} \cdot x_i}
\end{equation}
\begin{equation}
\label{eq:generalConstraint}
	\sum_{i=1}^{N}{S_{mi}}\cdot {x_i} = b_m  ~~~ m = 1 \cdots M
\end{equation}
where $N$ is the number of binary variables ($x_i$), $M$ is the number of constraints, $c_i$ and $b_m$ are real constants, and $S$ is a $M \cdot N$ matrix. The problem is transformed into a QUBO problem of the type:
\begin{equation}
\label{eq:qubo}
	min \ \bigg(\sum_{i=1}^{N}{c_{i} \cdot x_i} + A\sum_{m=1}^{M}\bigg[\sum_{i=1}^{N}{S_{mi}}\cdot {x_i} - b_m\bigg]^2 \bigg)
\end{equation}
In expression (\ref{eq:qubo}), the constraints are included in the objective function that we want to minimize. The quadratic form, and the value of a sufficiently large non-negative constant A, ensure that the minimum is obtained when all the constraints are matched. The specific setting of A will be explained in Section \ref{secModel}. After simplifying the quadratic terms ($x_i^2=x_i$ valid for binary variables), the QUBO problem becomes:

\begin{equation}
\label{eq:qubo}
	min \ \bigg(\sum_{i=1}^{N}{u_i}\cdot{x_i} + \sum_{i=1}^{N}\sum_{j=1}^{i}{v_{ij} \cdot x_i \cdot x_j}\bigg)
\end{equation}
where $u_i$ and $v_{ij}$ are real constants.

Then, with a substitution of variables, i.e.

\begin{equation}
\label{eq:substitution}    
x_i=\frac{1-z_i}{2}
\end{equation}
the problem is transformed into an Ising problem of the type:

\begin{equation}
\label{eq:ising}
	min \ \bigg( {\sum_{i=1}^{N}{h_i \cdot z_i} - \sum_{i=1}^{N}\sum_{j=1}^{i}{J_{ij} \cdot z_i \cdot z_j}} \bigg)
\end{equation}
where $h_i$ and $J_{ij}$ are real constants, and discrete variables $z_i$ can assume values +1 or -1, which correspond, respectively, to values 0 and 1 of the binary variables $x_i$.

Finally, a Hamiltonian operator is built with sums and tensor products of two basic one-qubit operators, the identity \textbf{I} and the Pauli operator \textbf{Z}, where, for each term in (\ref{eq:ising}), the operator $\textbf{Z}_i$ substitutes the variable $z_i$, and the identity operator $\textbf{I}$ is added for each variable $z$ that does not appear in the term. For example, with $N=4$ the term $z_2 \cdot z_3$ is substituted with $\textbf{I}_1 \otimes \textbf{Z}_2 \otimes \textbf{Z}_3 \otimes \textbf{I}_4$ or, more succinctly, $\textbf{I}_1 \textbf{Z}_2 \textbf{Z}_3 \textbf{I}_4$ or, even more briefly, $\textbf{Z}_2 \textbf{Z}_3$, where the identity operators are implicit. Moreover, the multiplications between the $z$ variables are substituted with the tensor products between the corresponding \textbf{Z} operators. With these rules, the Hamiltonian operator that corresponds to expression (\ref{eq:ising}) is:

\begin{equation}
\label{eq:IsingHamiltonian}
	\textbf{H} = {\sum_{i=1}^{N}{h_i \cdot \textbf{Z}_i} - \sum_{i=1}^{N}\sum_{j=1}^{i}{J_{ij} \cdot \textbf{Z}_i \otimes \textbf{Z}_j}}
\end{equation}

The operator \textbf{Z} is represented, in matrix notation, as:

\[
\textbf{Z} = \begin{bmatrix}
    1  & 0  \\
    0  & -1 \\
\end{bmatrix} .
\]
When the i-th qubit is measured with respect to the operator \textbf{Z}, denoted as $\textbf{Z}_i$, the measurement result is one of the two eigenvalues, $+1$ or $-1$, and the qubit collapses to the corresponding eigenvector (also called eigenstate), i.e., $[1,0]^T$ and $[0,1]^T$, respectively. In Dirac notation, the former eigenvector is denoted as $\ket{0}$, and the latter as $\ket{1}$.

Now, the problem is to find the minimum eigenvalue(s) of the operator (\ref{eq:IsingHamiltonian}). All these steps will be specialized in Section \ref{secModel}, with reference to the prosumer problem. Here, we would like to show, at least intuitively, for the reader who is not expert in quantum computing, why finding the minimum eigenvalue of the Hamiltonian operator corresponds to finding the setting of the $x$ variables that minimize the cost function $C$.



The Hamiltonian operator, thanks to the properties of the tensor product\footnote{The tensor product of two diagonal matrices is still diagonal.}, is always represented as a diagonal matrix, and the eigenstate corresponding to the minimum eigenvalue, i.e., the ground state, 
gives the solution to the problem. For example, let us assume that the problem to minimize, expressed in QUBO terms, is:

\begin{equation*}
\label{eq:qubo-ex}
	\ C = x_1 + 2 x_2 -4 x_1 x_2
\end{equation*}

The corresponding Ising formulation is 

\begin{equation*}
\label{eq:ising-ex}
	min \ \big(1/2 + z_1/2 - z_1 z_2 \big)
\end{equation*}

and the Hamiltonian of the problem is:

\begin{equation*}
\label{eq:Hamiltonian-ex}
	\textbf{H} = 1/2 \  (\textbf{I}_1 \otimes \textbf{I}_2) + 1/2\ (\textbf{Z}_1 \otimes \textbf{I}_2) - 
	(\textbf{Z}_1 \otimes \textbf{Z}_2)
\end{equation*}
The matrix notation of \textbf{H} is:
\[
\textbf{H} = \begin{bmatrix}
    0 & 0 & 0 & 0  \\
    0 & 2 & 0 & 0  \\
    0 & 0 & 1 & 0  \\
    0 & 0 & 0 & -1 \\
\end{bmatrix}
\]

In this case, the minimum eigenvalue is -1, which corresponds to the eigenvector $[0,0,0,1]^T$, which in turns corresponds to the basis state $\ket{11}$, since $\ket{11}$ = $\ket{1} \otimes \ket{1}$ = $[0,1]^T \otimes [0,1]^T$ = $[0,0,0,1]^T$. Therefore, the objective is to minimize the measurement result at the end of the circuit, since the minimum eigenvalue is achieved for the basis state $\ket{11}$ that represents the solution of the optimization problem, i.e., $x_1$=1 and $x_2$=1.  

As a final note, the output state prepared by the quantum circuit is in general the superposition of multiple basis states. The important thing is that, after applying the QAOA circuit, the amplitudes of the basis states that represent the solution become larger than the other amplitudes, so that a measurement provides a large probability of obtaining the desired solution.


\section{Solving the Prosumer Problem with QAOA}
\label{secModel}

In this section, the general steps described in the previous section are specialized for the case of the prosumer problem.
%
%
The prosumer problem can be formulated as in Eqs. (\ref{eq:generalObjFunc}) and (\ref{eq:generalConstraint}), as the problem of minimizing the energy cost incurred by the users of an energy community, while satisfying the constraints, i.e., addressing the user energy requirements and providing a feasible plan for each schedulable interruptible load.
A schedulable load is an electric load that can be shifted in a time interval, in accordance with the user preferences. A load is interruptible when can be stopped and resumed at a different time.
Some necessary definitions are provided in the following:
\begin{labeling}{xxxxx}
	\item [$L$] set of schedulable loads;
	\item [$\abs{L}$] number of schedulable loads;
	\item [$H$] set of scheduling hours;
	\item [$\abs{H}$] number of scheduling hours;
	\item [$\alpha_l;\beta_l$] start and end of time interval, defined by the user, for scheduling each load $l \in L$;
	\item [$\delta_l$] working time of the load $l \in L$, i.e., number of hours for which the load must be switched on;
	\item [$E_l$] power consumption (in Watts) of the load $l \in L$;
	\item [$E_{max}$] maximum nominal power of the user system;
	\item [$p^h$] cost of the electrical energy at the hour $h \in H$;
	\item [$x_l^h$] state of the load $l \in L$ at hour $h \in H$ ($1$ = on; $0$ = off).
\end{labeling}

Given the set of schedulable loads $L$ and the admissible scheduling interval $[\alpha_l;\beta_l]$ defined by the user, 
the problem is to minimize the following objective function: 
\begin{equation}
\label{eq:objFunc}
	\ C = {\sum_{l \in L}\sum_{h \in H}{(p^h \cdot x_l^h \cdot E_{l})}}
\end{equation}

Eq. (\ref{eq:objFunc}) aims to minimize the global energy cost, computed from the electrical energy tariffs $p^h$ and the power consumption of each load $E_l$, by scheduling the state (on or off) of the loads $x_l^h$. The following constraints need to be satisfied:
\begin{align}
\label{eq:constraintEmax}
	&\sum_{l\in L}({x_l^h} \cdot E_{l}) \leq E_{max} ~~~ \forall h\in H\\
\label{eq:constraintDelta}
	&\sum_{h=\alpha_l}^{\beta_l}{x_l^h}=\delta_l ~~~ \forall l \in L
\end{align}

In particular, inequalities (\ref{eq:constraintEmax}) force the value of the energy supplied to the loads, at each hour, to be lower or equal than the maximum nominal power available at the user's electrical system, $E_{max}$, while
equations (\ref{eq:constraintDelta}) ensure that each load $l$ is switched on exactly for $\delta_l$ hours inside the preference time interval $[\alpha_l;\beta_l]$. For this optimization problem, the number of binary variables $x_l^h$ is equal to $\sum_{l\in L}({\beta_l - \alpha_l})$ and the number of constraints is equal to $\abs{L} + \abs{H}$. It is noted that the values of energy, $E_{max}$ and $E_{l}$, are expressed as integers. This is not a limitation because, if fractional values are needed, they can be converted to integers through multiplication by an appropriate factor.

The inequalities (\ref{eq:constraintEmax}) are converted into equations, thus achieving the form of expression (\ref{eq:generalConstraint}),
by adding an extra integer variable $E_{res}^h$ that represents, for each hour, the residual energy that is not used, with $0 \le E_{res}^h \le E_{max}$:

\begin{equation}
\label{eq:conversioneInequToEqu}
    \sum_{l\in L}({x_l^h} \cdot E_{l}) + E_{res}^h = E_{max} ~~~ \forall h\in H\\
\end{equation}

The integer values of $E_{res}^h$ can be expressed through the use of $M$ slack binary variables, with $M = \lceil log(N) \rceil$, where $N$ is the number of integer values that $E_{res}^h$ can assume, with 
$N = E_{max} + 1$. With the use of the binary slack variables, the expression of $E_{res}^h$ becomes:  

\begin{equation}
\label{eq:conversioneIntToBin}
    E_{res}^h = \sum_{m = 1}^{M-1}(2^{m-1} \cdot y_{m}^h) + (N-2^{M-1}) \cdot y^h_{M}  ~~~ \forall h\in H
\end{equation}

In this way, the inequalities (\ref{eq:constraintEmax}) are converted into the equations:

\begin{multline}
    \sum_{l\in L}({x_l^h} \cdot E_{l}) + \sum_{m = 1}^{M-1}(2^{m-1} \cdot y_{m}^h) + (N-2^{M-1}) \cdot y^h_{M}\\ = E_{max} ~~~ \forall h\in H
\end{multline}

Now, the number of binary variables (including the original variables $x$ and the new slack variables $y$) becomes $\sum_{l\in L}({\beta_l - \alpha_l}) + M \cdot \abs{H}$. As we will see, each binary variable is represented as a qubit.


At this point, the ILP problem needs to be transformed into a QUBO problem. A function is built that incorporates the original objective function and a sum of penalties, each corresponding to a constraint. 
The addition of these penalties creates an augmented objective function to be minimized.
If the penalty terms can be driven to zero, through a proper setting of the binary variables, the augmented objective function becomes the original function to be minimized.
That is, the penalties equal zero for feasible solutions and equal some
positive penalty amount for infeasible solutions \cite{glover2022quantum}.  

The prosumer problem now becomes:

\begin{multline}
\label{eq:equation}
	min \bigg({\sum_{l \in L}\sum_{h \in H}{(p^h \cdot x_l^h \cdot E_{l})}}
	+A \cdot \bigg\{\sum_{h\in H}
	\\\bigg[\sum_{l\in L}({x_l^h} \cdot E_{l})
	+ \sum_{m = 1}^{M-1}(2^{m-1} \cdot y_{m}^h) + (N-2^{M-1}) \cdot y^h_{M} - E_{max}\bigg]^2\\
	+ \sum_{l\in L}\bigg[\sum_{h=\alpha_l}^{\beta_l}{x_l^h}-\delta_l\bigg]^2\bigg\}\bigg)
\end{multline}
where $A$ is a penalty coefficient computed as in Eq. \ref{eq:costanteA}, where $C_{up}$ and $C_{low}$ are, respectively, the upper and lower bounds of the cost function $C$, defined in Eqs. \ref{eq:costanteAUP} and \ref{eq:costanteALOW}. 

\begin{align}
    \label{eq:costanteA}
    &A = 1.0 + (C_{up} - C_{low})\\
    \label{eq:costanteAUP}
    &C_{up} = {\sum_{l \in L}\sum_{h \in H}{(p^h \cdot 1 \cdot E_{l})}}\\
    \label{eq:costanteALOW}
    &C_{low} = {\sum_{l \in L}\sum_{h \in H}{(p^h \cdot 0 \cdot E_{l})}}
\end{align}


With this setting, the value of $A$ is sufficiently large to ensure that the unconstrained problem of Eq. \ref{eq:equation} has the same optimal solution as the original constrained problem. This occurs because a possible penalty, deriving from the non-satisfaction of a constraint, cannot be compensated by the decreasing of the original objective function that is allowed by the missed constraint satisfaction.
%
As anticipated in Section \ref{secILP}, the QUBO expression can be simplified by noting that the square of a binary variable, $x^2$ (or $y^2$ in the case of a slack variable) is equal to $x$ (or $y$).

The last two steps are: (i) let every binary variable $x$ and $y$ undergo the substitution $x$ (or $y$) = $(1-z)/2$; (ii) build the Hamiltonian operator \textbf{H} by putting a $Z$ operator in place of each $z$ discrete variable. At this point, as discussed in Section \ref{secILP}, the problem becomes finding the minimum eigenvalue of the Hamiltonian operator. The qubits are prepared by QAOA to achieve with maximum probability the ground state, in which each qubit is represented by one of the basis states $\ket{0}$ or $\ket{1}$. Correspondingly, the solution to the problem is obtained by setting the binary variables (both the original ones and the slack variables) to $0$ or $1$.


\section{Example of Prosumer Problem}
\label{secExample}
This section illustrates an example of a prosumer problem for a single user of an energy community, equipped with two schedulable loads.
The preferential time interval is the same for both the loads, and is $[$09:00; 12:00$]$, with the possibility of switching on/off the loads at each hour.
The sets and constants are set as follows:
\begin{itemize}
    \item Set and number of schedulable loads:\\
	    $L$ = \{1,2\};\\
	    $\abs{L}$ = 2;
	\item Set and number of scheduling hours:\\
	    $H$ = \{1,2,3\};\\
	    $\abs{H}$ = 3;
	\item Time intervals and working times:\\
	    $[\alpha_1; \beta_1]$ = [09:00; 12:00];\\
	    $[\alpha_2; \beta_2]$ = [09:00; 12:00];\\
	    $\delta_1$ = 2 [h];\\
	    $\delta_2$ = 1 [h];
	
	\item Power consumption of the loads and nominal power of the system:\\
	    $E_1$ = 2 [kW];\\
	    $E_2$ = 1 [kW];\\
	    $E_{max}$ = 3 [kW];
	\item Hourly cost of the energy:\\
	    $p^1$ = 22 [\euro \ cent / kWh];\\
	    $p^2$ = 21 [\euro \ cent / kWh];\\
	    $p^3$ = 24 [\euro \ cent / kWh];
\end{itemize}

When replacing these values in Eq. \ref{eq:objFunc}, we obtain the following function to minimize:
\begin{multline}
\label{eq:objFuncExample}
		\ C = (
		     p^1 \cdot x_1^1 
	       + p^2 \cdot x_1^2 
	       + p^3 \cdot x_1^3) \cdot E_{1}\\
	       +(p^1 \cdot x_2^1 
	       + p^2 \cdot x_2^2 
	       + p^3 \cdot x_2^3) \cdot E_{2}
\end{multline}

The constraints expressed in Eq. \ref{eq:constraintEmax} and \ref{eq:constraintDelta} become:
\begin{align}
	&{x_1^1} \cdot E_{1} + {x_2^1} \cdot E_{2} \leq E_{max}\\
	&{x_1^2} \cdot E_{1} + {x_2^2} \cdot E_{2} \leq E_{max}\\
	&{x_1^3} \cdot E_{1} + {x_2^3} \cdot E_{2} \leq E_{max}\\
	&{x_1^1 + x_1^2 + x_1^3} = \delta_1\\
	&{x_2^1 + x_2^2 + x_2^3} = \delta_2
\end{align}

The number of binary variables is equal to $\sum_{l\in L}({\beta_l - \alpha_l}) = 6$, while the number of constraints is equal to $\abs{L} + \abs{H} = 5$.
As in Eq. \ref{eq:conversioneInequToEqu}, the inequality constraints are converted into equality constraints:
\begin{align}
	&{x_1^1} \cdot E_{1} + {x_2^1} \cdot E_{2} + E_{res}^{1} = E_{max}\\
	&{x_1^2} \cdot E_{1} + {x_2^2} \cdot E_{2} + E_{res}^{2} = E_{max}\\
	&{x_1^3} \cdot E_{1} + {x_2^3} \cdot E_{2} + E_{res}^{3} = E_{max}
\end{align}

At this point, we convert the integer variables $E_{res}^h$ into binary variables. Taking into account that these variables can assume only integer values included in the range $[0; E_{max}]$, we obtain that the parameters $N$ and $M$ of Eq. \ref{eq:conversioneIntToBin} are equal to 2 and 4, respectively. By substituting these values in the general equation, we obtain:
\begin{align}
    &E_{res}^{1} = y_1^1 + 2 \cdot y_2^1\\
    &E_{res}^{2} = y_1^2 + 2 \cdot y_2^2\\
    &E_{res}^{3} = y_1^3 + 2 \cdot y_2^3
\end{align}

The constraints can now be expressed in terms of the original load variables $x$ and the slack variables $y$, as follows:
\begin{align}
    &{x_1^1} \cdot E_{1} + {x_2^1} \cdot E_{2} + y_1^1 + 2 \cdot y_2^1 = E_{max}\\
	&{x_1^2} \cdot E_{1} + {x_2^2} \cdot E_{2} + y_1^2 + 2 \cdot y_2^2 = E_{max}\\
	&{x_1^3} \cdot E_{1} + {x_2^3} \cdot E_{2} + y_1^3 + 2 \cdot y_2^3 = E_{max}\\
	&{x_1^1 + x_1^2 + x_1^3} = \delta_1\\
	&{x_2^1 + x_2^2 + x_2^3} = \delta_2
\end{align}

The number of binary variables is equal to $\sum_{l\in L}({\beta_l - \alpha_l}) + M \cdot \abs{H} = 12$, where the first term is the number of the original variables and the second term is the number of slack variables.

The ILP problem is transformed into a QUBO problem, with the procedure discussed in Section \ref{secModel}, and the function to minimize becomes:
\begin{multline}
\label{eq:QUBOproblem}
	(
	   p^1 \cdot x_1^1 
	   + p^2 \cdot x_1^2 
	   + p^3 \cdot x_1^3) \cdot E_{1}\\
	   +(p^1 \cdot x_2^1 
	   + p^2 \cdot x_2^2 
	   + p^3 \cdot x_2^3) \cdot E_{2}\\
	+A \cdot \{
	 [{x_1^1} \cdot E_{1} + {x_2^1} \cdot E_{2} + y_1^1 + 2 \cdot y_2^1 - E_{max}]^2\\
	+[{x_1^2} \cdot E_{1} + {x_2^2} \cdot E_{2} + y_1^2 + 2 \cdot y_2^2 - E_{max}]^2\\
	+[{x_1^3} \cdot E_{1} + {x_2^3} \cdot E_{2} + y_1^3 + 2 \cdot y_2^3 - E_{max}]^2\\
	+[{x_1^1 + x_1^2 + x_1^3} - \delta_1]^2
	+[{x_2^1 + x_2^2 + x_2^3} - \delta_2]^2
	\}
\end{multline}
where the penalty coefficient $A$ is computed using Eq. \ref{eq:costanteA}, as follows: 

\begin{multline*}
    \label{eq:costanteAEsempio}
    A = 1.0 + \{44 \cdot 1 + 42 \cdot 1 + 48 \cdot 1 + 22 \cdot 1 + 21 \cdot 1 + 24 \cdot 1\}\\
    - \{44 \cdot 0 + 42 \cdot 0 + 48 \cdot 0 + 22 \cdot 0 + 21 \cdot 0 + 24 \cdot 0\}
\end{multline*}
and in this case $A$ is equal to 202.



As discussed before, the Hamiltonian operator is obtained with the substitution $x$ (or $y$) = $(1-z)/2$, and by putting the $Z$ operator in place of the $z$ discrete variables.

\begin{multline*}
  \textbf{H} = 79 \cdot Z_1
+ 80 \cdot Z_2
+ 77 \cdot Z_3
- 112 \cdot Z_4
- 111.5 \cdot Z_5
- 113 \cdot Z_6\\
+ 101 \cdot Z_1 Z_2
+ 101 \cdot Z_1 Z_3
+ 101 \cdot Z_2 Z_3
+ 202 \cdot Z_1 Z_4\\
+ 202 \cdot Z_2 Z_5
+ 101 \cdot Z_4 Z_5
+ 202 \cdot Z_3 Z_6
+ 101 \cdot Z_4 Z_6\\
+ 101 \cdot Z_5 Z_6
+ 202 \cdot Z_1 Z_7
+ 101 \cdot Z_4 Z_7
+ 404 \cdot Z_1 Z_8\\
+ 202 \cdot Z_4 Z_8
+ 202 \cdot Z_7 Z_8
+ 202 \cdot Z_2 Z_9
+ 101 \cdot Z_5 Z_9\\
+ 404 \cdot Z_2 Z_{10}
+ 202 \cdot Z_5 Z_{10}
+ 202 \cdot Z_9 Z_{10}
+ 202 \cdot Z_3 Z_{11}\\
+ 101 \cdot Z_6 Z_{11}
+ 404 \cdot Z_3 Z_{12}
+ 202 \cdot Z_6 Z_{12}
+ 202 \cdot Z_{11} Z_{12}\\
+ 2019.5
\end{multline*}

The QAOA algorithm can now be executed to achieve the ground state of \textbf{H}, which encodes the values of the binary variables that minimize the QUBO function (\ref{eq:QUBOproblem}).






\section{Results}
\label{secResults}
In this section, we first discuss some results obtained by solving the simple prosumer problem formulated in Section \ref{secExample}, which regards two loads that can be scheduled in a time interval of three hours. Then, we extend the same problem to a larger number of hours, which of course increases the needed number of qubits. 


We adopted the quantum simulator provided by the IBM Quantum Experience, and the Qiskit library, which enables the use of a common programming language, i.e., Python, to formulate quantum algorithms.
The problem has 9 feasible solutions, all of which are discovered by the QAOA algorithm. For each solution, Table \ref{tab:probability} reports the values of the six binary variables, and the corresponding value of the cost function, obtained from Eq. \ref{eq:objFuncExample}.
 \begin{table}[!h]
 \caption{Feasible solutions and values of the cost function for the problem of Section \ref{secExample}.}
\label{tab:probability}
\centering
\begin{tabular}{|c|c|c|c|c|c|c|c|}
\hline
\begin{tabular}[c]{c}feasible\\ solution
\end{tabular}
& $x_1^1$ & $x_1^2$ & $x_1^3$ & $x_2^1$ & $x_2^2$ & $x_2^3$ &
\begin{tabular}[c]{c}cost \\ {[}\euro\ cent{]}
\end{tabular}    \\ \hline
1                                                              & 1                       & 1                       & 0                       & 0                       & 1                       & 0                       & 107      \\ \hline
2                                                              & 1                       & 1                       & 0                       & 1                       & 0                       & 0                       & 108      \\ \hline
3                                                              & 1                       & 1                       & 0                       & 0                       & 0                       & 1                       & 110      \\ \hline
4                                                              & 0                       & 1                       & 1                       & 0                       & 1                       & 0                       & 111      \\ \hline
5                                                              & 0                       & 1                       & 1                       & 1                       & 0                       & 0                       & 112      \\ \hline
6                                                              & 1                       & 0                       & 1                       & 0                       & 1                       & 0                       & 113      \\ \hline
7                                                              & 0                       & 1                       & 1                       & 0                       & 0                       & 1                       & 114      \\ \hline
8                                                              & 1                       & 0                       & 1                       & 1                       & 0                       & 0                       & 114      \\ \hline
9                                                              & 1                       & 0                       & 1                       & 0                       & 0                       & 1                       & 116      \\ \hline
\end{tabular}
\end{table}

The best solution is obtained when the first load, with a power consumption of $2\ kW$, is scheduled from 09:00 to 11:00, and the second load, with a power consumption of $1\ kW$ , is scheduled from 10:00 to 11:00. The corresponding global energy cost is 1.07 \euro, computed as reported in Eq. (36).
\begin{multline}
		\ C = (
		     0.22 \text{\euro} / kW \cdot 1 
	       + 0.21 \text{\euro} / kW \cdot 1 
	       + 0.24 \text{\euro} / kW \cdot 0) \cdot 2 kW\\
	       +(0.22 \text{\euro} / kW \cdot 0 
	       + 0.21 \text{\euro} / kW \cdot 1 
	       + 0.24 \text{\euro} / kW \cdot 0) \cdot 1 kW
\end{multline}


The number of binary variables, and therefore of qubits, is proportional to the number of time slots $\abs{H}$ that are admissible for the scheduling, as reported in Table \ref{tab:numBinaryVar}. For example, if 5 hours are considered, the problem includes 10 load variables and 10 slack binary variables used to manage the inequalities, therefore the overall number of qubits is 20.
\begin{table}[!h]
\caption{Number of binary+slack variables, for three cases with different time intervals.}
\label{tab:numBinaryVar}
\centering
\begin{tabular}{|c|c|c|c|}
\hline
hours            & 3   & 4   & 5     \\ \hline
binary variables & 6+6 & 8+8 & 10+10 \\ \hline
\end{tabular}
\end{table}

The quantum computers that are built by the largest IT companies currently provide tens or even hundreds of qubits, and even more powerful computers are expected in the coming years. Theoretically, a large number of qubits enables the solution of very complex prosumer problems, which would not be manageable with a classical computer. Noise and decoherence can limit the full exploitation of the quantum computing power, but notable improvements are expected also in this aspect.


In this paper, we only report the solutions obtained with a simulator, while current and future efforts are devoted to the test of real quantum computers. The execution on simulators is very useful in the first phase, to check the correctness of solutions found by QAOA, and then to compare simulations with experiments on the quantum hardware. However, the use of a simulator limits the number of qubits. 
Indeed, a quantum register with n qubits lives in a Hilbert space with dimensionality $2^n$, and an operator -- in our case the Hamiltonian operator -- is represented as a $2^n$ x $2^n$ matrix. Hence, the simulator encounters memory problems very soon as the number of qubits increases. Table \ref{tab:executionTime} reports the average execution time experienced with three, four, and five scheduling hours, for different values of \textit{reps}, which is the number of times that the QAOA quantum gates are repeated\footnote{It is known that more repetitions help to improve the solution.}. As expected, the execution time increases with the number of qubits and with the number of repetitions, and remains acceptable, in the order of seconds. However, with 5 schedulable hours, and 20 qubits, the quantum simulator outputs an error regarding the matrix size, for the reason explained before.


\begin{table}[!h]
\caption{Execution time (s), for the case of three, four, and five scheduling hours.}
\label{tab:executionTime}
\centering
\begin{tabular}{ccccc}
\multicolumn{2}{c}{\multirow{2}{*}{}}                               & \multicolumn{3}{c}{hours}                                                                           \\ \cline{3-5} 
\multicolumn{2}{c}{}                                                & \multicolumn{1}{|c|}{3}      & \multicolumn{1}{c|}{4}      & \multicolumn{1}{c|}{5}                  \\ \cline{2-5} 
\multicolumn{1}{c|}{\multirow{5}{*}{reps}} & \multicolumn{1}{c|}{1} & \multicolumn{1}{c|}{9.765}  & \multicolumn{1}{c|}{22.905} & \multicolumn{1}{c|}{\multirow{5}{*}{-}} \\ \cline{2-4}
\multicolumn{1}{c|}{}                      & \multicolumn{1}{c|}{2} & \multicolumn{1}{c|}{19.532} & \multicolumn{1}{c|}{33.302} & \multicolumn{1}{c|}{}                   \\ \cline{2-4}
\multicolumn{1}{c|}{}                      & \multicolumn{1}{c|}{3} & \multicolumn{1}{c|}{23.748} & \multicolumn{1}{c|}{52.497} & \multicolumn{1}{c|}{}                   \\ \cline{2-4}
\multicolumn{1}{c|}{}                      & \multicolumn{1}{c|}{4} & \multicolumn{1}{c|}{32.700} & \multicolumn{1}{c|}{60.002} & \multicolumn{1}{c|}{}                   \\ \cline{2-4}
\multicolumn{1}{c|}{}                      & \multicolumn{1}{c|}{5} & \multicolumn{1}{c|}{35.159} & \multicolumn{1}{c|}{88.539} & \multicolumn{1}{c|}{}                   \\ \cline{2-5} 
\end{tabular}
\end{table}


\section{Conclusions}
\label{secConclusions}
This paper addresses the solution of a typical prosumer problem with a quantum computing approach. We described the steps through which the problem can be formulated as a binary optimization problem, and then transformed into the problem of finding the ground state of a diagonal Hamiltonian operator, which is the kind of problem that can be solved by the Quantum Approximate Optimization Algorithm (QAOA). 
We solved a simple prosumer problem using the IBM quantum simulator. The results confirm the ability of the simulator to find the best solution at minimum cost, but the problem becomes unfeasible as the number of qubits increases, due to the size of the matrix that encodes the quantum Hamiltonian operator used to solve the problem. Current work is devoted to the execution of the QAOA algorithm on a real quantum platform, and on the assessment of the results for larger prosumer problems.


\balance



\bibliographystyle{plain} 
\bibliography{bibliography}


\end{document}